\documentclass[letterpaper, 10 pt, conference]{ieeeconf}  

\IEEEoverridecommandlockouts                              
\usepackage{graphics}
\usepackage{epsfig}
\usepackage{amsmath}
\usepackage{amssymb}
\usepackage{setspace}
\usepackage{psfrag}
\usepackage{cite}
\usepackage{epstopdf}

\newfont{\smalll}{cmr8}

\def\span{\mathrm{span}}

\def\IC{\hbox{C\hskip-
.5em\raise.5ex\hbox{$\scriptscriptstyle\mid$}}\ }
\def\Ic{\hbox{\smalll C\hskip-
.5em\raise.3ex\hbox{$\scriptscriptstyle\mid$}}\ }

\def\T={\buildrel {\scriptscriptstyle\triangle} \over =}

\def\sqr#1#2{{\vcenter{\vbox{\hrule height.#2pt\hbox{\vrule
width.#2pt height#1pt \kern#1pt\vrule width.#2pt}\hrule
height.#2pt}}}}

\def\block-diag{\mathop{\rm block{\scriptstyle -}diag}}

\def\pmbb#1{\setbox0=\hbox{#1}\raise 0.5ex\box0}

\newcommand{\bequ}{\begin{eqnarray}}
\newcommand{\eequ}{\end{eqnarray}}

\newcommand{\rom}{\mathrm}

\newcommand {\beq}      {\begin{equation}}
\newcommand {\eeq}      {\end{equation}}

\def\IC{{\mathbb C}}

\renewcommand\theenumi{\roman{enumi}}
\renewcommand\theenumii{\alph{enumii}}
\renewcommand\theenumiii{\arabic{enumiii}}
\renewcommand\theenumiv{\Alph{enumiv}}

\def\sc{s_{\rm c}}

\def\sc{s_{\rm c}}

\title{\LARGE \bf
Scalability Concept for Predictable Closed-Loop Response \protect\\ of Adaptive Controllers
}

\author{Simon P.  Schatz and Tansel Yucelen
\thanks{Simon P. Schatz is a Graduate Research Assistant at the Institute of Flight System Dynamics, Technische Universit\"at M\"unchen, 85748 Garching, Germany,
        {\tt\small simon.p.schatz@tum.de}}%
\thanks{Tansel Yucelen is an Assistant Professor at the Department of Mechanical and Aerospace Engineering, Missouri University of Science and Technology,
        Rolla, MO 65409, USA,
        {\tt\small tyucelen@mst.edu}}%
}

\newtheorem{remark2}{Remark}

\begin{document}

\maketitle
\thispagestyle{empty}
\pagestyle{empty}

\begin{abstract}

We introduce a new concept called \textit{scalability} to adaptive control in this paper.
In particular, we analyze how to scale learning rates of adaptive weight update laws of various adaptive control schemes with respect to given command profiles to achieve a predictable closed-loop response. 
An illustrative numerical example is provided to demonstrate the proposed concept, which emphasize that it can be an effective tool for validation and verification of adaptive controllers.

\end{abstract}

\section{SCALABLE PERFORMANCE IN MODEL REFERENCE ADAPTIVE CONTROL}
\label{sec:MRAC}
In this section, scalability is shown in the standard model reference adaptive control (MRAC) architecture.
\subsection{MRAC Problem Formulation}
 Consider the uncertain dynamical system given by
\bequ
	\dot{x}(t) = Ax(t)+B\Lambda u(t) + B \Delta(x(t)), \quad x(0)=x_0, \label{eqn:MRAC_1}
\eequ
where $x(t)\in\Re^n$ is the accessible state vector, $u(t) \in \Re^m$ is the control input vector, $\Delta(x(t)): \Re^n \rightarrow \Re^m$ is an \textit{uncertainty},
$A \in \Re^{n \times n}$ is a known system matrix,
$\Lambda \in \Re^{m \times m}_{+}$ is an \textit{unknown} control effectiveness matrix,
and $B \in \Re^{n \times m}$ is a known control input matrix. We assume that the pair $(A,B)$ is controllable. Additionally, we assume
\bequ
	\Delta(x(t)) = \Lambda \begin{bmatrix} W_\rom{x}^\rom{T} & W_\rom{c}^\rom{T} & w_\kappa\end{bmatrix} \omega(t), \label{eqn:MRAC_2}
\eequ
where $W_\rom{x} \in \Re^{n \times m}$ represents an uncertainty in the system matrix, $W_\rom{c} \in \Re^{l \times m}$ represents an uncertainty in the command input matrix, $\omega = \begin{pmatrix}x(t)^\rom{T} & c(t)^\rom{T}& \kappa\end{pmatrix} \in \Re^{n+l+1}$ is a known regressor vector, $c(t)\in\Re^l$ is the uniformly continuous bounded command, $\kappa$ is a constant, and $w_\kappa \in \Re^{m}$ represents a constant disturbance.
The reference system is given by
\bequ
	\dot{x}_r(t) = A_rx_r(t)+B_rc(t), \quad x_r(0)=x_{r0},  \label{eqn:MRAC_3}
\eequ
where $x_r(t)\in\Re^n$ is the reference model state vector, $A_r \in \Re^{n \times n}$ is the desired Hurwitz system matrix, and $B_r \in \Re^{n \times l}$ is the command input matrix.
The control signal $u(t)$ is given as
\bequ
	u(t) = -K_xx(t) + K_cc(t) - u_\rom{ad}(t), \label{eqn:MRAC_4}
\eequ
where $u_\rom{ad}(t) \in \Re^{m}$ is the adaptive control input,
$K_x \in \Re^{m \times n}$ is the nominal feedback matrix and $K_c \in \Re^{m \times l}$  is the nominal feedforward matrix chosen such that $A - BK_x = A_r$ and $BK_c = B_r$.
Using (\ref{eqn:MRAC_2}) and (\ref{eqn:MRAC_4}) in (\ref{eqn:MRAC_1}), yields
\bequ
	\dot{x}(t) = A_rx(t)+B_rc(t)+B\Lambda W^\rom{T}\omega(t) - B\Lambda u_\rom{ad}(t), \label{eqn:MRAC_6}
\eequ
where
\bequ
	W \triangleq \begin{bmatrix} W_\rom{x}^\rom{T}-\Lambda^{*}K_x&W_\rom{c}^\rom{T}+\Lambda^{*}K_c&w_\kappa\end{bmatrix}^\rom{T}, \label{eqn:MRAC_7}
\eequ
and $\Lambda^{*} \triangleq \bigl[I_m - \Lambda^{-1}\bigl]$. We use the adaptive control law
\bequ
	u_\rom{ad}(t) = \hat{W}^\rom{T} (t)\omega(t), \label{eqn:MRAC_8}
\eequ
where $\hat{W}(t)\in \Re^{(n+l+1) \times m}$ is the adaptive weight matrix satisfying the adaptive weight update law
\bequ
	 \dot{\hat{W}}(t) &=& \Gamma \omega(t)e^\rom{T}(t)PB, \quad \hat{W}(0) = \hat{W}_{0}, \label{eqn:MRAC_9}
\eequ
$e(t) \triangleq x(t) - x_r(t)$ is the tracking error, and $P\in\Re^{n \times n}$ is the positive definite solution of the Lyapunov equation
\bequ
	 Q+A_r^\mathrm{T}P+PA_r=0, \label{eqn:LYAP}
\eequ
where $Q\in\Re^{n \times n}$ is a positive definite design matrix. Finally, the uncertain dynamical system (\ref{eqn:MRAC_1}) can now be given as 
\bequ
	\dot{x}(t) = A_rx(t)+B_rc(t)-B\Lambda \tilde{W}^\rom{T}(t)\omega(t), \, x(0)=x_0, \label{eqn:MRAC_10}
\eequ
where $\tilde{W}(t) \triangleq \hat{W}(t) - W \in \Re^{(n+l+1) \times m}$ is the adaptive weight estimation error.

\subsection{Scalability}
\label{MRAC_scale}
Now, we assume that the control engineer has found an appropriate adaptive control performance for a certain command history $c_{0}(t)$ and a specified learning rate $\Gamma_{0}$, resulting in the adaptive weight update law
\bequ
	 \dot{\hat{W}}(t) = \Gamma_{0} \omega(t)e^\rom{T}(t)PB, \quad \hat{W}(0) = \hat{W}_{0}, \label{eqn:SCL_1}
\eequ
For any scaled command profiles $c(t)=\alpha c_0(t)$ with scalar scaling command coefficients $\alpha\neq0$ given a Lyapunov design matrix Q  it is possible to achieve scaled system responses by choosing $\Gamma = \Gamma_{0} / \alpha^{2}$.
To show this, we define $z(t)\triangleq x(t)/\alpha$, $z_0\triangleq x_0/\alpha$, $z_\rom{r}(t)\triangleq x_\rom{r}(t)/\alpha$, $z_\rom{r0}\triangleq x_\rom{r0}/\alpha$, $e_\rom{z}(t) \triangleq z(t) - z_\rom{r}(t) = e(t)/\alpha$, $\kappa = \alpha$ and
\bequ
	\omega_z(t) \triangleq \alpha \omega(t) = \alpha \begin{pmatrix} z(t) & c_0(t) & 1 \end{pmatrix}^\mathrm{T}. \label{eqn:SCL_2}
\eequ
By applying this transformation to the uncertain dynamical system (\ref{eqn:MRAC_10}) and the weight update law (\ref{eqn:MRAC_9}), we have
\bequ
	\dot{z}(t) = A_rz(t)+B_rc_0(t)-B\Lambda \tilde{W}^\rom{T}\omega_z(t), \, z(0)=z_0,\label{eqn:SCL_3}\\
	\dot{z}_r(t) = A_rz(t)+B_rc_0(t), \quad z_r(0)=z_{r0}, \quad \quad  \quad \quad \quad \label{eqn:SCL_4}\\
	\dot{\hat{W}} (t)= \Gamma_{0} \omega_z(t)e^\rom{T}_{z}(t)PB, \quad \hat{W}(0) = \hat{W}_{0}. \quad \quad \quad \quad  \quad \label{eqn:SCL_5}
\eequ
Note that the equations (\ref{eqn:SCL_3}), (\ref{eqn:SCL_4}), and (\ref{eqn:SCL_5}) hold for any $\alpha \neq 0$. Further, note that the uncertain system (\ref{eqn:MRAC_10},\ref{eqn:SCL_3}) and the reference system (\ref{eqn:MRAC_3},\ref{eqn:SCL_4}) are scalable in the sense that state histories can be given by a nominal system response scaled by $\alpha$.  

\section{OTHER MRAC SCHEMES}
\label{sec:EXT}
The scalability notion is applicable to all MRAC based schemes under the assumption that the states are applicable.
In particular, in this section it is shown that the $\sigma$-modification and $e$-modification adaptive control architectures {\cite{sigma:1}},{\cite{emod:1}}, frequency-limited adaptive controllers{\cite{yucelen:2}}, adaptive control architectures employing closed-loop reference models {\cite{notable:3,notable:4}}, and command governor-based adaptive controllers {\cite{yucelen:cg}} can all be modified in order to achieve predictable performances as shown previously. 

\subsection{$\sigma$ and $e$ modification architectures}
\label{sec:sigma}
These robustness modifications have been introduced in order to avoid the phenomena of parameter drift and increase the robustness with respect to unmodeled dynamics. The architectures modify the adaptive weight update law by augmenting it with a ``damping-like" term.

In {\cite{sigma:1}} the standard MRAC adaptive weight update law was modified as
\bequ
	 \dot{\hat{W}}(t) = \Gamma \omega(t)e^\mathrm{T}(t)PB - \sigma \hat{W}(t), \quad \hat{W}(0) = \hat{W}_{0}, \label{eqn:EXT_1}
\eequ
where $\sigma>0$ is a damping coefficient used to "pull" the estimated adaptive weights towards the origin. It was claimed that this $\sigma$-modification prevented the estimated adaptive weight from becoming unbounded.  
By introducing a scaling factor, as done in Section \ref{MRAC_scale}, the $\sigma$-modifed adaptive weight update law is given as 
\bequ
	\dot{\hat{W}} (t)= \Gamma_{0} \omega_z(t)e^\mathrm{T}_{z}(t)PB - \sigma \hat{W}(t), \quad \hat{W}(0) = \hat{W}_{0}. \label{eqn:EXT_2}
\eequ
where $\Gamma=\Gamma_0/\alpha^{2}$, $\omega_z(t)$ and $e_z(t)$ are defined in the previous section. It can be readily seen that the adaptive weight response, as before, is invariant to the scaling factor $\alpha$. Furthermore, scalability of the system states and inputs is also evident since the reference system (\ref{eqn:MRAC_3})
and the uncertain system (\ref{eqn:MRAC_10}) are not modified.

In {\cite{emod:1}} the standard MRAC adaptive law was further modified by replacing $\sigma$ in (\ref{eqn:EXT_1}) with a time-varying damping coefficient given by $\sigma_e\left\Vert e(t) \right\Vert_2$. Therefore, the effect of the modification was determined by the norm of the system's tracking error. The so-called $e$-modification adaptive weight update law is given as
\bequ
	 \dot{\hat{W}} (t)= \Gamma \omega(t)e^\mathrm{T}(t)PB - \sigma_e \left\Vert e(t) \right\Vert_2 \hat{W}(t), \label{eqn:EXT_3}
\eequ
$\hat{W}(0) = \hat{W}_{0}$, where $\sigma_e>0$. Similar to (\ref{eqn:EXT_1}), by introducing a scaling factor (\ref{eqn:EXT_3}) can be rewritten as 
\bequ
	 \dot{\hat{W}} (t)= \Gamma_{0} \omega_z(t)e_z^\mathrm{T}(t)PB - \sigma_{0} \left\Vert e_z(t) \right\Vert_2 \hat{W}(t), \label{eqn:EXT_4}
\eequ
$\hat{W}(0) = \hat{W}_{0}$, where $\sigma_e=\sigma_0/\alpha$ and, as before, $\Gamma=\Gamma_0/\alpha^{2}$. Note that, as seen for the $\sigma$-modification case, (\ref{eqn:EXT_3}) is invariant to $\alpha$ and, therefore, scalability results.     


It should be noted that all the adaptive control architectures considered in this section are obtained with simple augmentations of the standard MRAC adaptive weight update law. In general, if the augmentation is invariant to the scaling factor $\alpha$ then the modified adaptive control framework will be scalable in the sense introduced in this paper.

\subsection{Frequency-Limited Adaptive Control}
The frequency limited adaptive control architecture introduced in {\cite{yucelen:2}} employs a gradient based modification term and a low pass filter. It is claimed that the modification term filters high-frequency content out of
the adaptive weight update law, allowing for the controller to be tuned with high learning rates in order to enable robust and fast adaptation. The adaptive weight update law is given by
\bequ
	\dot{\hat{W}} (t)= \Gamma\omega(t)e^\mathrm{T}(t)PB - \sigma [\hat{W}(t) - \hat{W}_f(t)], \label{eqn:FL_1}
\eequ
$\hat{W}(0) = \hat{W}_{0}$, where $\sigma>0$ is a modification gain and $\hat{W}_f(t) \in \Re^{(n+l+1) \times m}$ is the low-pass filtered weight estimate of $\hat{W}(t)$, satisfying
\bequ
	\dot{\hat{W}}_f (t)= \Gamma_f [\hat{W}(t) - \hat{W}_f(t)], \; \hat{W}_f(0) = \hat{W}_{0}, \label{eqn:FL_2}
\eequ
where $\Gamma_f \in \Re^{(n+l+1) \times (n+l+1)}$ is a positive definite filter gain matrix such that $\lambda_\mathrm{max}(\Gamma_f) \leq \gamma_\mathrm{f,max}$ and $\gamma_\mathrm{f,max}>0$ is a design parameter.

The adaptive weight update law (\ref{eqn:FL_2}) can incorporate the scaling factor $\alpha$ as
\bequ
	\dot{\hat{W}}(t) = \Gamma_0\omega_z(t)e_z^\mathrm{T}(t)PB - \sigma [\hat{W}(t) - \hat{W}_f(t)], \label{eqn:FL_3}
\eequ
$\hat{W}(0) = \hat{W}_{0}$, where $\Gamma=\Gamma_0/\alpha^2$, $e_z(t) = e(t)/\alpha$, and $\omega_z(t)=\omega(t)/\alpha$. Note that once again the adaptive weight update law is invariant with respect to the scaling factor $\alpha$.
Therefore, as discussed in the previous section, it can be concluded that a system employing this adaptive control framework will have predictably scalable responses.

\subsection{Reference Model Modification}
\label{clrm}
In {\cite{notable:3,notable:4}} the reference model was modified by feeding back the tracking error in order to improve the transient performance of MRAC controllers.
Therefore, the uncertain dynamical system (\ref{eqn:MRAC_10}) and the adaptive weight update law (\ref{eqn:MRAC_9}) are not changed and can be scaled as shown in Section \ref{MRAC_scale}.
However, the reference model is given by
\bequ
	\dot{x}_r(t) = A_rx_r(t)+B_rc(t)+Le(t), \quad x_r(0)=x_{r0},  \label{eqn:SP_1}
\eequ
where $L \in \Re^{n\times n}$ is a positive definite matrix. The scaling factor can then be introduced to the modified reference model by employing, as before, the relations $z_r(t) = x_r(t)/\alpha$, $z_{r0} = x_{r0}/\alpha$, $e_z(t) = e(t)/\alpha$, and $c(t) = \alpha c_0(t)$, resulting in
\bequ
	\dot{z}_r(t) = A_rz_r(t)+B_rc_0(t)+Le_z(t), \quad z_r(0)=z_{r0}.  \label{eqn:SP_2}
\eequ
Hence, scalability for adaptive control architectures with modified reference models is obtained.

\subsection{Command Governor Adaptive Control}
\label{sec:cg_frame}
Here, the scalability notion is applied to the command governor framework for adaptive control{\cite{yucelen:cg}}.
%

\addtolength{\textheight}{-6.6cm}

Hence, the overall command is given by
\bequ
	c(t) \triangleq c_D(t) + c_g(t),  \label{eqn:CG_1}
\eequ
where $c_D(t) \in \Re^{m}$ is the bounded, desired tracking command (the original $c(t)$ from the sections above). The additional command $c_g(t) \triangleq K_c^{-1}\left[B^\mathrm{T}B\right]^{-1}B^\mathrm{T}g(t) \in \Re^{m}$, $\mathrm{det}(K_c) \neq 0$ is based on a linear system, which is defined as
\bequ
	\dot{\xi}(t) &=& -\lambda \xi(t) + \lambda e(t), \quad \xi(0) = 0,  \label{eqn:CG_2} \\
	g(t) &=& \lambda \xi(t) + \left[A_r - \lambda I_n \right] e(t), \label{eqn:CG_3}
\eequ
where $\xi(t) \in \Re^{n}$ denotes the command governor states, $g(t) \in \Re^{n}$ is the command governor output, and $\lambda>0$ is the command governor gain. Since the additional command is applied on both reference model and nominal controller, the error dynamics of the system do not change and therefore, we have
\bequ
	    \dot{e}(t) = A_r e(t) - B\Lambda \tilde{W}^\rom{T}(t)\omega(t), \quad e(0) = x_0 -x_{r0},   \label{eqn:CG_4}
\eequ
which can be written as
\bequ
	   \Lambda \tilde{W}^\rom{T}(t)\omega(t) = [B^\mathrm{T}B]^{-1}B^\mathrm{T}\{A_re(t) - \dot{e}(t)\}.  \label{eqn:CG_5}
\eequ
Applying (\ref{eqn:CG_1}), (\ref{eqn:CG_2}), (\ref{eqn:CG_3}), and (\ref{eqn:CG_5}) onto the uncertain system dynamics (\ref{eqn:MRAC_10}) using $G=B\left[B^\mathrm{T}B\right]^{-1}B^\mathrm{T}$, we have
\bequ
	  \dot{x}(t) = A_rx(t)+B_rc_D(t)+G\{ \lambda \xi(t)- \lambda e(t)-\dot{e}(t)\},  \label{eqn:CG_6}
\eequ
$x(0) = x_0$. In {\cite{yucelen:cg}} it is shown that $\lambda \xi(t)- \lambda e(t)-\dot{e}(t) = 0$ for $\lambda\rightarrow\infty$ and that the overall system is stable.
\begin{remark2}
Although the reference model is modified, the closed loop uncertain system still tracks the desired reference model given by
\bequ
	  \dot{x}_{r,D}(t) = A_rx_{r,D}(t)+B_rc_D(t), \quad x_{r,D}(0)=x_{r0}  \label{eqn:CG_7}
\eequ
as the last term of (\ref{eqn:CG_6}) is appoximately zero for large $\lambda$.
\end{remark2}
\begin{remark2}
The command governor gain $\lambda$ can be used to determine a trade off between command governor and adaptive control.
Furthermore, note that no adaptive control would be necessary
for $\lambda\rightarrow\infty$, which is of no practical relevance.
For more information about the command governor refer to {\cite{yucelen:cg}}.
\end{remark2}

Now, considering scalability, assume there was a $c_0(t)$ with a certain reference performance and a learning rate $\Gamma_0$. Then, applying a command profile $c_D(t)=\alpha c_0(t)$ and a scaled adaptive gain $\Gamma = \Gamma_0/\alpha^2$, scalability can be achieved. Using $\xi_z(t)=\xi(t)/\alpha$, $g_z(t)=g(t)/\alpha$, and $e_z(t)=e(t)/\alpha$, we have
\bequ
	\dot{\xi}_z(t) &=& -\lambda \xi_z(t) + \lambda e_z(t), \quad \xi_z(0)=0  \label{eqn:CG_8} \\
	g_z(t) &=& \lambda \xi_z(t) + \left[A_r - \lambda I_n \right] e_z(t). \label{eqn:CG_9}
\eequ
Hence, $c_{g,z}(t) = \alpha c_g(t)$ and $c(t) = \alpha c_0 + \alpha c_{g}$ holds, which implies that the reference model is also scalable as shown in Section \ref{MRAC_scale}. The transformed uncertain system dynamics are given by (using $f_z(t)\triangleq \lambda \xi_z(t)- \lambda e_z(t)-\dot{e}_z(t)$)
\bequ
	  \dot{z}(t) = A_rz(t)+B_rc_{0}(t)+Gf_z(t), \; z(0) = z_0, \label{eqn:CG_10}
\eequ
which shows scalability of the uncertain system's dynamics. Additionally, the invariance of the adaptive weight update law (\ref{eqn:SCL_5}) to the scaling factor stays untouched. Consequently, the scalability approach introduced in this paper also holds for the command governor framework.


\begin{thebibliography}{99}
%
%
%
%
%
%
%

\bibitem{yucelen:2}
T. Yucelen and W. M. Haddad, ``Low-frequency learning and fast adaptation in model reference adaptive control,'' \textit{IEEE Transactions on Automatic Control}, 2013.



%
%
%

\bibitem{yucelen:cg}
T. Yucelen and E. N. Johnson, ``A new command governor architecture for transient response shaping,'' \textit{International Journal of Adaptive Control and Signal Processing}, 2013.

\bibitem{dydek}
Z. Dydek, H. Jain, J. Jang, A. M. Annaswamy, and E. Lavretsky, ``Theoretically Verifiable Stability Margins for an Adaptive Controller,'' \textit{AIAA Guidance, Navigation, and Control Conference}, 2006

\bibitem{sigma:1}
P. Ioannou and P. Kokotovic, ``Instability analysis and improvement of robustness of adaptive control,'' \textit{Automatica}, vol. 20, pp. 583--594, 1984.

\bibitem{emod:1}
K. S. Narendra and A. M. Annaswamy, ``A new adaptive law for robust adaptation without persistent excitation,'' \textit{IEEE Transactions on Automatic Control}, vol. 32, pp. 134--145, 1987.


\bibitem{notable:3}
E. Lavretsky, ``Reference dynamics modification in adaptive controllers for improved transient performance,'' \textit{AIAA Guidance, Navigation, and Control Conference}, 2011.

\bibitem{notable:4}
T. E. Gibson, A. M. Annaswamy, and E. Lavretsky, ``Adaptive systems with closed-loop reference models: Stability, robustness, and transient performance,'' \textit{IEEE Transactions on Automatic Control} (submitted).


\end{thebibliography}
\end{document}